\documentclass[twocolumn,pra,aps,superscriptaddress,showpacs]{revtex4-1}
\usepackage{mathptmx}
\usepackage{subfigure}
\usepackage{psfrag,graphicx}
\usepackage{pict2e}
\usepackage{dcolumn}
\usepackage{amsmath,amssymb}
\usepackage{bm}
\usepackage{color}
\usepackage{latexsym}
\usepackage{epstopdf}
\usepackage{color}
\usepackage[english]{babel}
\usepackage{amsfonts}
\usepackage{bm}
\usepackage{natbib}

\usepackage{appendix}
\usepackage{bbold}
\usepackage{placeins}
\usepackage{appendix}
\definecolor{mygrey}{gray}{0.35}
\definecolor{myblue}{rgb}{0.2,0.2,0.8}
\definecolor{myzard}{cmyk}{0,0,0.05,0}
\definecolor{mywhite}{rgb}{1,1,1}
\definecolor{mywhite}{rgb}{1,1,1}
\definecolor{myred}{rgb}{1,0.,0.3}
\usepackage[colorlinks=true,citecolor=myblue,linkcolor=myred]{hyperref}
\def\be{\begin{equation}}
\def\ee{\end{equation}}
\def\ba{\begin{align}}
\def\enda{\end{align}}
\def\bi{\begin{itemize}}
\def\ei{\end{itemize}}

\def\adag{{a^{\dag}}}

\begin{document}
\title{Quantum sensing close to a dissipative phase transition: symmetry breaking and criticality as metrological resources}

\author{Samuel Fern\'andez-Lorenzo}
\email{S.Fernandez-Lorenzo@sussex.ac.uk}
\affiliation{Department of Physics and Astronomy, University of Sussex, Falmer, Brighton BN1 9QH, UK}

\author{Diego Porras}
\email{D.Porras@sussex.ac.uk}
\affiliation{Department of Physics and Astronomy, University of Sussex, Falmer, Brighton BN1 9QH, UK}

\date{\today}

\begin{abstract}
We study the performance of a single qubit-laser as a quantum sensor to measure the amplitude and phase of a driving field. 
By using parameter estimation theory we show that certain suitable field quadratures are optimal observables in the lasing phase. The quantum Fisher information scales linearly with the number of bosons and thus the precision can be enhanced by increasing the incoherent pumping acting on the qubit. 
If we restrict ourselves to measurements of the boson number observable, then the optimal operating point is the critical point of the lasing phase transition.
Our results point out to an intimate connection between symmetry breaking, dissipative phase transitions and efficient parameter estimation.
\end{abstract}


\maketitle

\section{Introduction}
Quantum sensing and metrology is likely to be a key practical application of quantum technologies. 
It has been established both by theory and experiments that quantum effects can be exploited to increase the accuracy of measurement devices 
\cite{caves81prd,braunstein94prl,sanders95prl,bollinger96,giovannetti11natphys}. 
Practical applications, however, face significant challenges.
In an ideal scenario quantum metrology requires the preparation of many-particle entangled states by quantum operations that so far are only possible with a few degrees of freedom. 
Dissipation and noise pose severe limitations which  often hinder the metrological advantages of entangled states \cite{huelga97prl,escher11,guta12natcomm,chin12prl,huelga16prl}.
Quantum setups such as superconducting circuits \cite{majer07,houck12natphys,devoret13sci} and trapped ions \cite{Schneider12rpp,blatt12natphys} offer us the opportunity to engineer quantum states of matter with a high degree of control over interactions and dissipation. 
It has been shown that dissipation may be actually exploited as an effective tool in quantum state engineering \cite{verstraete09,barreiro11}. 
The question naturally arises, whether we can use dissipation to design metrological protocols and sensors \cite{dallatore13prl,gonzaleztudela13prl}. 
We propose two working principles for such quantum sensors. First, one could exploit the sensitivity of a dissipative steady-state to an external field which explicitly breaks some suitable underlying symmetry. The second route could take advantage of the sensitivity at the critical point of a dissipative phase transition\cite{Zanardi08,wang14njp,salvatori14pra,guta16pra}. Such a sensor would have the advantage that state preparation is not required and, furthermore, dissipation is a control parameter of the sensor dynamics, rather than an error source. 

\begin{figure}[h!]
  \includegraphics[width=0.37\textwidth]{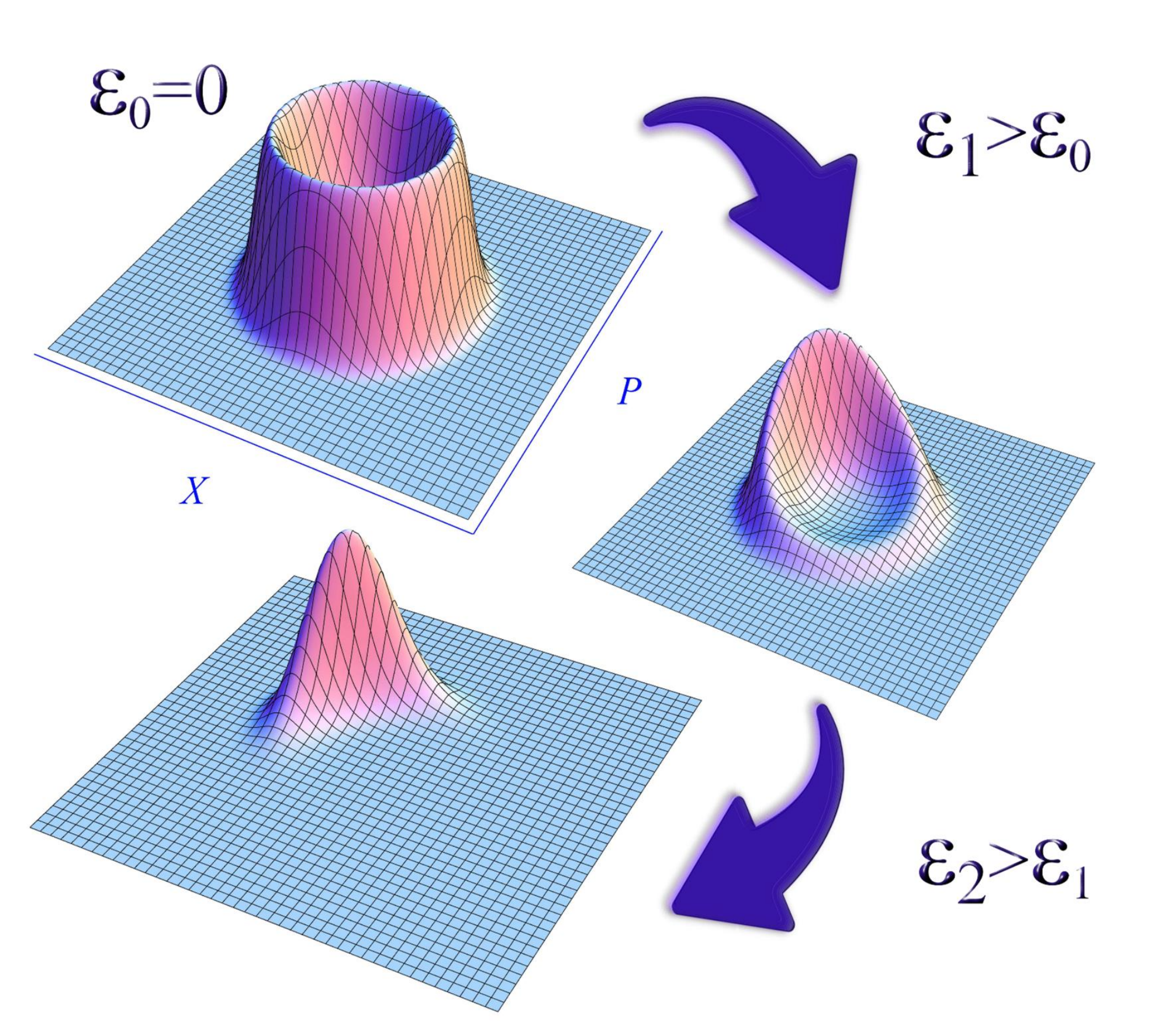}
  \caption{Explicit symmetry breaking of the \textit{Glauber-Sudarshan P} representation of the lasing steady state \eqref{Ppolar} when an external periodic driving field $\epsilon\neq0$ is introduced.}
  \label{fig:fig0} 
\end{figure}

In this work, we show that a single qubit laser is a minimalist model where both ideas can be tested.
%
%
A macroscopic laser with $n$ photons can be described by a coherent state of the light field with a mean value $\langle a \rangle = \sqrt{n} e^{i \theta}$, which assumes the spontaneous breaking of the underlying lasing phase symmetry by choosing an arbitrary value of $\theta$ \cite{molmer97pra}. This approach can be justified by assuming an infinitesimal field (e.g. an environmental fluctuation) that fixes the laser phase \cite{scully70pra}. However, in a finite size system (e.g. a single qubit laser) an external field with finite amplitude, $\epsilon$, is required to explicitly break the phase symmetry (see Fig.\ref{fig:fig0}). 
Here the thermodynamic limit is found when $n\rightarrow\infty$ \cite{hwang15}, at which the system undergoes a spontaneous symmetry breaking, i.e. $\lim_{\epsilon\to 0}\lim_{n\to\infty} \langle a \rangle \neq 0$. This relation implies that the order parameter $\langle a \rangle$ must increase with the system size $n$ with an scaling yet to be determined, leading to a high sensitivity to $\epsilon$.

Our article is structured as follows. Firstly, we present a semi-classical description in phase space of a single-qubit laser in the presence of a weak symmetry breaking driving field. This allows us to estimate analytically the quantum Fisher information related to the amplitude $|\epsilon|$ and phase $\phi$ of the driving, which further shows the connection between symmetry breaking and efficient parameter estimation. We identify the optimal observables that fully exploit the system metrological capacity. Non-equilibrium criticality is then examined as an alternative metrological resource with non-optimal protocols using the average number of bosons. We conclude with a discussion of possible error sources as well as applications. 

\section{Single qubit laser}
In this work we consider a bosonic mode coupled  by a Jaynes-Cummings interaction to a two-level system (qubit) with levels $|g\rangle$ and $|e\rangle$. Additionally, we introduce a periodic driving which becomes the target weak field. Both the qubit and driving frequencies are assumed to be resonant with the bosonic mode. In an interaction picture rotating at the mode frequency, the coherent dynamics is described by the Hamiltonian
\begin{eqnarray}
H &=& H_{\rm JC}+H_{\rm d} , \label{Hamiltonian} \nonumber
\\
H_{\rm JC} &=& g(\sigma^+ a+ \adag\sigma^-), \quad H_{\rm d}=  
\epsilon^* a  + \epsilon a^{\dag} ,
\end{eqnarray}
where $\epsilon = |\epsilon| e^{i \phi}$, with $|\epsilon|$ and $\phi$ being the driving amplitude and phase, respectively.  $\sigma^{\pm}$ are the ladder operators of the two-level system, $\sigma^{+}=|e\rangle \langle g|$ and $\sigma^{-}=|g\rangle \langle e|$. In addition to this coherent dynamics, the system is subjected to incoherent pumping of the qubit and losses of the bosonic mode with rates $\gamma$ and $\kappa$, respectively. The resulting dissipative process is well captured by the following master equation for the system density matrix $\rho$,
\begin{equation}
\dot{\rho}=-i [H,\rho]+\mathcal{L}_{\{\sigma^+ , \gamma\}}(\rho) + 
\mathcal{L}_{\{a , \kappa\}}(\rho),
\label{Liouvillian}
\end{equation}
where Lindblad super-operators are defined as $\mathcal{L}_{\{O , \Gamma\}}(\rho)= \Gamma (2 O\rho O^{\dag}-O^{\dag}O\rho - \rho O^{\dag}O)$.  
In a mean field approximation to the case without driving ($\epsilon  = 0$) the steady-state is determined by the \textit{pump parameter}, $C_{\rm p} \equiv g^2/\gamma\kappa$. This sets a dissipative phase transition into a lasing phase when $C_{\rm p} > 1$ \cite{BreuerPet}, being $\langle a \rangle$ the order parameter.
To evaluate the response of the single-qubit laser to an external driving, we need to go beyond mean-field theory. Since we are only interested in the output laser field, we start by finding an effective Liouvillian able to describe the reduced dynamics of the bosonic mode. This can be accomplished in a strong pumping regime  \cite{Sargent74,Mandel95}, \textit{i.e} $\gamma \gg \kappa, g, |\epsilon|$, in which
the qubit can be adiabatically eliminated, leading to an effective quartic master equation for the bosonic mode (see App.\ref{App:AppendixA} for a detailed derivation), 
\begin{align}
	\dot{\rho}_f&=-i[\epsilon^* a  + \epsilon a^{\dag},\rho_f] + \mathcal{L}_{\{\adag, A\}}(\rho_f)+\mathcal{L}_{\{a, C\}}(\rho_f)+\label{finalrhof} \nonumber  \\ 
	&+ \mathcal{L}_{\{a\adag, B\}}(\rho_f)-\mathcal{L}_{\{\adag^2, B\}}(\rho_f).	
\end{align}
We have defined the coefficients 
$A=g^2/\gamma$, 
$B=2g^4/\gamma^3$, 
$C=\kappa$, and $\rho_f = {\rm Tr}_{\rm quibt}{\{\mathcal{L}(\rho)\}}$ is the reduced density matrix of the bosonic field. Our expression is valid in a regime of strong incoherent pumping, such that the probability of occupation of the ground state can be neglected. This condition is justified both below the lasing phase transition, $C_{\rm p} < 1$, or slightly above the threshold, $C_{\rm p} \gtrsim 1$ (see App.\ref{App:AppendixA}). \\

\section{Semi-classical limit}  
The master equation obtained in \eqref{finalrhof} is still challenging to be tackled analytically. By using phase space methods we shall obtain a Fokker-Planck equation valid in a regime with high number of bosons \cite{Gardiner,Haken75}.
This will allow us to get analytical results that will be assessed below by comparing to exact numerical calculations.
We start by introducing the \textit{coherent state} or \textit{ Glauber-Sudarshan P representation} of the effective master equation \cite{Carmichael,Mandel95}, defined as 
\begin{equation}
\rho(t)=\int d^2\alpha P(\alpha,\alpha^*,t) |\alpha\rangle\langle \alpha |, \label{coherentP}
\end{equation}
where $|\alpha\rangle$ is the coherent state $|\alpha\rangle=\exp{(\alpha a^{\dag}-\alpha^* a)}|0\rangle$. The function $P(\alpha,\alpha^*)$ plays a role analogous to that of a classical probability distribution over $|\alpha\rangle\langle \alpha |$, with the normalization condition $\int d^2\alpha P(\alpha,\alpha^*,t) = 1$, 
and expectation values of normal ordered operators, $\langle (\adag)^p a^q \rangle=\int d^2\alpha (\alpha^*)^p \alpha^q P(\alpha,\alpha^*)$. Note that $P$ is actually a quasi-probability distribution, since it is in general not a positive distribution function. \

By substituting the representation \eqref{coherentP} of $\rho$ into Eq. \eqref{finalrhof}, one may convert the operator master equation into an equation of motion for $P(\alpha,\alpha^*,t)$. This can be accomplished by using the following equivalences,
\begin{align}
	a|\alpha\rangle\langle \alpha |&=\alpha |\alpha\rangle\langle \alpha | \\ 
	|\alpha\rangle\langle \alpha | \adag &=\alpha^* |\alpha\rangle\langle \alpha | \\
	\adag |\alpha\rangle\langle \alpha | &=\left( \frac{\partial}{\partial\alpha} +\alpha^* \right)  |\alpha\rangle\langle \alpha | \label{alphaderiv}\\ 
	|\alpha\rangle\langle \alpha |a &=\left( \frac{\partial}{\partial\alpha^*} +\alpha \right)  |\alpha\rangle\langle \alpha | . \label{alphaderiv2}
\end{align}
An integration by parts with the assumption of zero boundary conditions at infinity, which introduces an extra minus sign for each differential operator, converts the integrand of \eqref{coherentP} into a product of $ |\alpha\rangle\langle \alpha |$ and a \textit{c}-number function of $\alpha,\alpha^*$. This leads to a differential equation for $P(\alpha,\alpha^*,t)$. When the laser is operating near the steady state and above threshold, $|\alpha|^2$ is a large number of the order of the average number of photons. Notice also that $B$ is a very small coefficient compared to $A$, such that $B/A\propto(g/\gamma)^2 \ll 1$. Consequently, we shall retain only the most important terms in $B$. This corresponds to dropping any contribution smaller than $B|\alpha|^2 \alpha$. In doing so, we end up with the following Fokker-Planck equation for $P$,
\begin{equation}
\frac{\partial P}{\partial t}= -\frac{\partial}{\partial \alpha} [(A-C-B|\alpha|^2)\alpha  -\epsilon'] P +c.c. +2A\frac{\partial^2P}{\partial \alpha\partial\alpha^*} \label{PFokker}
\end{equation}
where $\epsilon'\equiv i\epsilon$. Let us write this equation in cartesian coordinates, with $\alpha=x_1+i x_2$ and $\partial/\partial\alpha=1/2(\partial/\partial x_1- i\partial/\partial x_2)$ then
\begin{equation}
\frac{\partial P}{\partial t}= -\sum^2_{i=1}\frac{\partial}{\partial x_i} [(A-C-B \vec{x}^2)x_i  -\epsilon'_i] P +\frac{A}{2}\sum^2_{i=1}\frac{\partial^2P}{\partial x^2_i} \label{fokker}
\end{equation}
where we introduce the two-dimensional vectors $\vec{x}=(x_1,x_2)$ and $\vec{\epsilon'}=(\Re(\epsilon'),\Im(\epsilon'))$. In the stationary state $\partial P/\partial t=0$, Eq.\eqref{fokker} may be rewritten as $\sum_i\partial J_i/\partial x_i=0$, where the current $\vec{J}$ is defined by 
\begin{equation}
J_i=  [(A-C-B \vec{x}^2)x_i  -\epsilon'_i]-\frac{A}{2}\frac{\partial P}{\partial x_i}.
\end{equation}
When the drift vector $A_i\equiv [(A-C-B \vec{x}^2)x_i  -\epsilon'_i]$ satisfies the potential condition $\partial A_i/\partial x_j=\partial A_j/\partial x_i$, as it does in our case, the solution to the Fokker-Planck equation is derived by imposing $\vec{J}=0$ \cite{Mandel95}. This leads to a differential equation for $P$ that can be directly integrated to give
\begin{equation}
P(\vec{x})=  \frac{1}{N} \exp{\left\{\frac{1}{A}\left[\left(A-C-\frac{B}{2}\vec{x}^2\right)\vec{x}^2-2\vec{\epsilon'}\cdot\vec{x} \right] \right\} }, \label{Pest}
\end{equation}
where $N$ is a normalization constant. The steady-state solution \eqref{Pest} can be conveniently expressed in polar coordinates $\alpha=re^{i\theta}$ as follows,
\begin{equation}
P(r,\theta)=  \frac{1}{N} \exp{(-\lambda r^4+ \mu r^2 -2\nu r\sin{(\theta-\phi)} )} , \label{Ppolar}
\end{equation}
where we have introduced the parameters 
$\lambda = B/2A$,
$\mu=(A-C)/A$, and $\nu=|\epsilon|/A$. Note that the probability distribution \eqref{Ppolar} is positive, which indicates that the steady-state admits a classical description.
Eq. \eqref{Ppolar} can be used to calculate expectation values in the steady state through parametric derivatives of the normalization constant, $N$. In the App.\ref{App:AppendixC}, it is detailed how to approximately calculate $N$ using the Laplace's method together with explicit expressions of useful observables.
In the absence of driving ($\epsilon=0$), the laser phase is uniformly distributed in $[0,2\pi]$, implying that any average field quadrature vanishes. In contrast, when $\epsilon\neq0$, the driving field explicitly breaks the phase symmetry and the state adopts a preferred phase with exponential sensitivity as illustrated in Fig.\ref{fig:fig0}. According to Eq.\eqref{Ppolar} we expect the output laser field to have a phase delay of $\pi/2$ with respect to the input driving.

\begin{figure}[h!]
			\includegraphics[width=0.45\textwidth]{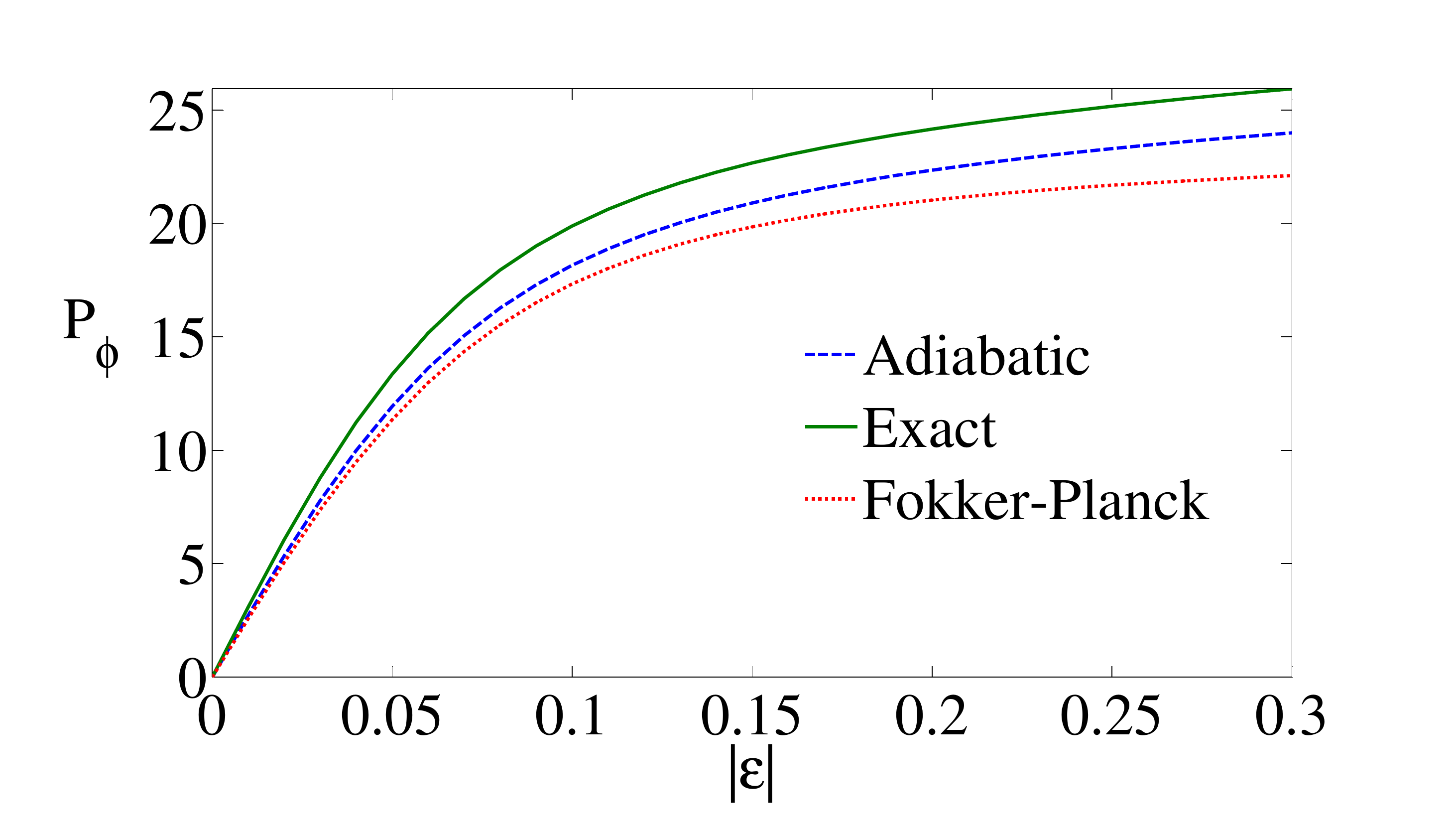}
  \caption{Plot of the averaged field quadrature $\langle \hat{P}_\phi \rangle$ as a function of the amplitude $|\epsilon|$, showing the comparison of exact calculations of Eq.\eqref{Liouvillian} (solid line), the adiabatic equation \eqref{finalrhof} (dashed line) and analytical result by the Fokker-Planck equation (dotted line). Parameters: $g=58,\gamma=3000,\kappa=1$.}
  \label{fig:fig1}
\end{figure}

Not any explicit symmetry breaking may lead to an advantageous sensing scheme. However, when this is associated to a spontaneous symmetry breaking (SSB) in the thermodynamic limit, the corresponding order parameter is expected to be very sensitive to such symmetry breaking field. Such subclass of non-trivial explicit symmetry breaking process is henceforth referred as \textsl{induced symmetry breaking}. 
In our case, this general symmetry argument is translated as a high sensitivity of the coherent component $\langle a \rangle$ to $\epsilon$. The average field quadrature $\langle \hat{P}_\phi \rangle=\langle i(ae^{-i\phi}-a^{\dag}e^{i\phi}) \rangle$ will be shown to be particularly sensitive to the external driving, analytically given by
\begin{equation}
\langle \hat{P}_\phi \rangle=2r_0\frac{I_1(2\nu r_0)}{I_0(2\nu r_0)}
\underset{\nu r_0 \ll 1}{\approx} \frac{2 r_0^2}{A}|\epsilon| ,
\label{quadrature}
\end{equation}
where $I_n(z)$ are the modified Bessel functions of the first kind, and $r_0^2$ stands for steady average number of bosons with no driving (see App.\ref{App:AppendixC} for detailed derivation),
\begin{equation}
\langle n\rangle_{\epsilon \approx 0} = r_0^2 = (A-C)/B .
\label{nph}
\end{equation}
The SSB of the lasing phase transition here implies $\lim_{\epsilon\to 0}\lim_{r_0^2\to\infty} \langle \hat{P}_\phi \rangle \neq 0$. This entails a certain scaling of $\langle \hat{P}_\phi \rangle$ with the system size, here $r_0^2$, now explicitly given by \eqref{quadrature}.
%
Figure \ref{fig:fig1} shows the comparison of these results with numerical calculations of the 
exact and the adiabatic equation, Eqs.\eqref{Liouvillian} and \eqref{finalrhof} respectively. 
From Fig.\ref{fig:fig1} we differentiate two distinct regimes. First, a linear regime $\langle \hat{P}_\phi \rangle \propto |\epsilon|$ if $|\epsilon|$ is small enough, where $\langle \hat{P}_\phi \rangle$ scales linearly with the number of bosons. Essentially, the more pronounced the slope is, the higher the sensor sensitivity will be. Second, a saturation regime where $\langle \hat{P}_\phi \rangle^2 \approx \langle a^\dagger a \rangle $ and the laser admits a fully classical description \cite{Mandel95}. \\

%
\section{Quantum Fisher information and optimal measurements}
Eq. \eqref{quadrature} suggests that the induced symmetry breaking allows us to measure weak field amplitudes $|\epsilon|\ll 1 $. 
The capability of this sensing scheme will be mainly determined by its resolution. The theory of quantum Fisher information \cite{Dowling15,Demkow14} provides us with an ultimate lower bound on the precision of parameter estimation that is possible in a quantum model, which will be used to assess the maximum metrological capacity of the single-qubit laser.

Assume that a target parameter $\varphi$ is encoded in a certain density matrix $\rho_\varphi$. The quantum \textit{Cramer-Rao bound} establishes a lower bound to the error in the estimation of $\varphi$,
\begin{equation}
\Delta^2\varphi\geq \frac{1}{N_{\rm exp}F_Q[\rho_\varphi]} \label{CramerRao}
\end{equation}
where $F_Q[\rho_\varphi]$ is the quantum Fisher information (QFI) and $N_{\rm exp}$ the number of experiment repetitions. 
The QFI can be viewed as a quantitative measure of distinguishability of a state $\rho_\varphi$ from its neighbors $\rho_{\varphi+\delta\varphi}$. Thus it can be used as a quantitative characterization of the maximal sensor resolution. A measurement scheme that saturates the bound Eq.\eqref{CramerRao} is called optimal. The \textit{symmetric logarithmic derivative} operation (SLD) is known to be optimal for all quantum states \cite{Braunstein96}. It is defined by the Hermitian operator $L_\varphi$ satisfying the relation
\begin{equation}
 \partial_\varphi \rho_\varphi=\frac{1}{2}(\rho_\varphi L_\varphi+L_\varphi \rho_\varphi)\label{SLD}.
\end{equation}
The QFI is then given by $F_Q[\rho_\varphi]=Tr\left\{\rho_\varphi L_\varphi^2\right\}$. In the eigenbasis of $\rho_\varphi=\sum_i {\lambda_i(\varphi)|e_i(\varphi)\rangle\langle e_i(\varphi)|}$, the SLD is written as
\begin{equation}
 L_\varphi[\rho_\varphi]=\sum_{\substack{i,j \\\lambda_i+\lambda_j\neq 0}} \frac{2\langle e_i(\varphi)|\dot{\rho}_\varphi|e_j(\varphi)\rangle}{\lambda_i(\varphi)+\lambda_j(\varphi)}|e_i(\varphi)\rangle\langle e_j(\varphi)|\label{SLDnumeric},
\end{equation}

First we shall focus on the estimation of the field amplitude $|\epsilon|$ for a given known phase $\phi$. By using the analytical result for the steady state \eqref{Ppolar}, we aim for deriving theoretical results for the SLD as well as the QFI. To do so, it is necessary to solve the operator equation \eqref{SLD} for $L_{|\epsilon|}$. In this context, a comprehensive solution of Eq.\eqref{SLD} is already known for Gaussian states in phase space, \textit{i.e} quadratic in $\alpha,\alpha^*$ \cite{Monras13}. 
Assuming the adiabatic elimination regime, \textit{i.e} $\gamma \gg \kappa, g, |\epsilon|$, the coefficients  $A,B$ satisfy $A/B\propto (\gamma/g)^2 \gg 1$. Hence the $P$ function \eqref{Ppolar} can be well approximated by the following Gaussian-like approximation,
\begin{equation}
P(r,\theta)=  N^{-1} \exp{(-\frac{(r-r_0)^2}{2\sigma^2}- \nu r\sin{(\theta-\phi)} )} \label{PpolarGauss}
\end{equation}
where $r^2_0=\mu/(2\lambda)$ and $\sigma^2=1/(4\mu)$. Even though this represents a simplification with respect to the original $P$ function \eqref{Ppolar}, the state is still not Gaussian in the variables $\alpha,\alpha^*$, for which exact solutions are known for the SLD and QFI \cite{Monras13}. Even so, let us try to solve the equation \eqref{SLD} in the coherent state representation. Using \eqref{PpolarGauss}, the l.h.s of the equation \eqref{SLD} gives
\begin{equation}
\partial_{|\epsilon|}P(r,\theta)=  \left(-N^{-1}\partial_{|\epsilon|}N+\frac{i}{A}(\alpha e^{-i\phi}-\alpha^*e^{i\phi})\right)P \label{derivP}.
\end{equation}
It turns out that $N^{-1}\partial_{|\epsilon|}N$ is equivalent to the average of the field quadrature  $\langle \hat{P}_\phi \rangle=\langle i(ae^{-i\phi}-a^{\dag}e^{i\phi}) \rangle$. This result induces us to introduce the ansatz $L_{|\epsilon|}=S_0+S a+S^*a^{\dag}$, with $S_0,S$ proper coefficients, which corresponds essentially to the measurement of a suitable field quadrature. Inserting this ansatz in the r.h.s of \eqref{SLD} and bearing in mind the equivalences (\ref{alphaderiv},\ref{alphaderiv2}), we have
\begin{equation}
L_{|\epsilon|}\rho=\int_{0}^\infty\int_{0}^{2\pi} rd\theta dr(S_0+S\alpha+S^*(\alpha^*-\partial_\alpha))P \label{ansatz1}
\end{equation}
with analogous result for $\rho L_{|\epsilon|}$. In a deep lasing regime (well above threshold but still within the validity regime of \eqref{finalrhof}) where $r_0 \gg \sigma$, the derivative $\partial_\alpha$ in Eq.\eqref{ansatz1} can be simplified assuming that $\alpha=re^{i\theta}\approx r_0e^{i\theta}$, yielding
\begin{multline} \label{ansatz}
\partial_\alpha P=\frac{e^{-i\theta}}{2}(\frac{\partial}{\partial r}-\frac{i}{r}\frac{\partial}{\partial\theta})P= \\
=(-\frac{\alpha^*}{2\sigma^2}+\frac{r_0}{2\sigma^2}e^{-i\theta}+\frac{i|\epsilon|}{A}e^{-i\phi})P\approx  \frac{i|\epsilon|}{A}e^{-i\phi}P .
\end{multline}
Identifying now terms from both sides of the equation \eqref{SLD}, the SLD reads
\begin{equation}
 L_{\epsilon}[\rho_{|\epsilon|}]=\frac{1}{A}\left(-\langle \hat{P}_\phi\rangle +\frac{|\epsilon|}{A}+ \hat{P}_\phi \right) .
\end{equation}
 The contribution $|\epsilon|/A^2$ can be neglected in comparison with the contribution given by $\hat{P}_\phi$, leading to the SLD $ L_{\epsilon}[\rho_{|\epsilon|}]=\left(-\langle \hat{P}_\phi\rangle+ \hat{P}_\phi \right)/A $. Happily, this in turn implies that $\langle L_{|\epsilon|} \rangle=0$, a property that any SLD must fulfill according to its own definition \eqref{SLD}. 
The QFI may now be calculated as $F_Q[\rho_\varphi]=Tr\left\{\rho_\varphi L_\varphi^2\right\}$ in terms of a parametric derivative of the normalization constant $N$ introduced in \eqref{Ppolar}, specifically as the fluctuations of $\hat{P}_\phi$ (see App.\ref{App:AppendixC}),
\begin{equation}
F_Q[\rho_{|\epsilon|}]= \frac{2r_0^2}{A^2}\left( 1+ \frac{I_2(2\nu r_0)}{I_0(2\nu r_0)}- 2\left(\frac{I_1(2\nu r_0)}{I_0(2\nu r_0)}\right)^2  \right)\underset{\nu r_0 \ll 1}{\approx} \frac{2 r_0^2}{A^2}. \label{Fisher}
\end{equation}

\begin{figure}[h!]
  \includegraphics[width=0.45\textwidth]{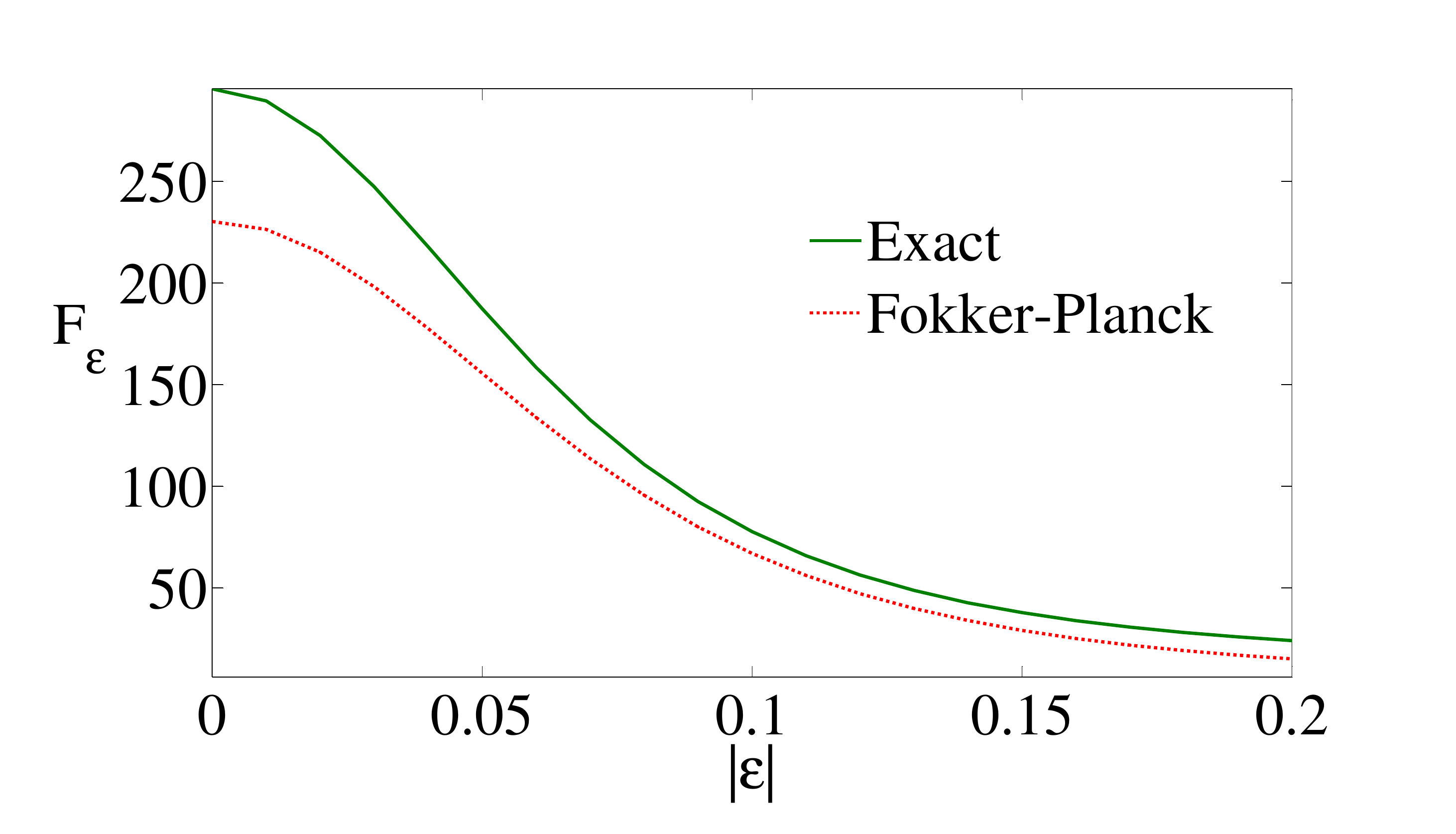}
  \caption{Plot of the quantum Fisher information $F_{|\epsilon|}$ as a function of the amplitude $|\epsilon|$ that shows the comparison between an exact calculation given by Eq.\eqref{Liouvillian} and the analytical result \eqref{Fisher}. Parameters: $g=58,\gamma=3000,\kappa=1$. }
  \label{fig:fig2} 
\end{figure}

In Fig. \ref{fig:fig2} we show a comparison between the analytical result \eqref{Fisher} and an exact numerical calculation of Eq.\eqref{Liouvillian} by using \eqref{SLDnumeric}. 
There are two important conclusions that are drawn from \eqref{Fisher}. Firstly, it shows that the metrological capacity for estimating $|\epsilon|$ is maximal when the induced symmetry breaking occurs, and decreases as the symmetry is already broken. This is intuitively natural since the parameter $|\epsilon|$ is directly associated with the symmetry breaking, and the gain of information is maximal at that point. This feature can be reasonably expected in any sensing scheme relying on spontaneous symmetry breaking as this one. Consequently, this type of sensing is advantageous when measuring extremely weak fields as the precision naturally increases in such domain. The parameters of the laser can be adjusted so that the amplitude remains in the first order approximation, where the precision remains constant for a fixed amplitude as Eq.\eqref{Fisher} indicates. Secondly, the QFI scales linearly with the steady average number of bosons $n$ as $|\epsilon|\rightarrow 0$. In the macroscopic limit, defined here as $r_0^2 \to \infty $, $F_Q$ diverges as a result of the sensitivity of the steady-state to an infinitesimal perturbation, giving rise to a spontaneous symmetry breaking. These results show a useful connection between symmetry breaking and efficient parameter estimation. 
The prior knowledge of $\phi$ in estimating $|\epsilon|$ may be eluded by performing an average of different quadratures over the range $[0, 2\pi]$, decreasing the 
QFI by a $1/2$ factor but still conserving the same scaling.

In the light of these results, we examine whether a similar approach can be used for measuring the phase $\phi$ for a given amplitude. 
A completely analogous procedure can now be used to solve again the operator equation \eqref{SLD}. Now the l.h.s of the equation \eqref{SLD} gives
\begin{equation}
\partial_{\phi}P(r,\theta)=  \left(-N^{-1}\partial_{\phi}N+2\nu(\alpha e^{-i\phi}+\alpha^*e^{i\phi})\right)P \label{derivP2}.
\end{equation}
The term $N^{-1}\partial_{\phi}N=\langle 2\nu r \cos(\theta -\phi) \rangle$ is easily shown to be zero.
Using then a linear ansatz $L_{\phi}=S_0+S a+S^*a^{\dag}$, the r.h.s of \eqref{SLD} is analogous to Eq.\eqref{ansatz1}. The comparison between both sides of the equation yields the SLD,
\begin{equation}
 L_{\phi}[\rho_{\phi}]=\nu\left( \hat{X}_{\phi} \right).
\end{equation}
where $\hat{X}_\phi$ is the field quadrature $\hat{X}_\phi = (a e^{-i\phi}+a^{\dag}e^{i\phi})$. The operator $L_{\phi}[\rho_{\phi}]$ also satisfies $\langle L_{\phi} \rangle =0$ as required by the definition \eqref{SLD} .
The QFI is then $F_Q[\rho_{\phi}]= Tr\left\{\rho_\phi L_\phi^2\right\}=\nu^2\langle \hat{X}^2_{\phi} \rangle$, which turns out to be equivalent to
\begin{equation}
F_Q[\rho_{\phi}]= \nu\langle \hat{P}_{\phi} \rangle 
\underset{\nu r_0 \ll 1}\approx \frac{2 r_0^2 |\epsilon|^2}{A^2} ,\label{FisherPhase} 
\end{equation}
where we have used Eq.\eqref{quadrature}. 
This result predicts that the QFI scales linearly with $n$ and quadratically with the field amplitude as $|\epsilon|\rightarrow 0$. Graphically, the behavior of the QFI in this case is indirectly given in Fig.\ref{fig:fig1}. In contrast to Eq.\eqref{Fisher} for estimating the amplitude, here the QFI increases with $|\epsilon|$ since naturally a non-zero signal is required to have a localized phase.
Note that the optimal observable $\hat{X}_{\phi}$ depends itself on the target parameter, $\phi$. To operate in the optimal measurement regime we need a first estimation of the observable, $\phi_0$. If such estimation satisfies the condition $\delta\phi=(\phi-\phi_0)\ll 1$, the quadrature $\hat{X}_{\phi_0}$ leads to an optimal protocol for estimating $\phi$, with a precision determined by Eq. (\ref{FisherPhase}). This requirement is analogous to the optimal free precession time in Ramsey spectroscopy \cite{Ludlow15}. 

In summary, our optimal scheme makes use of the coherent component $\langle a \rangle$ to estimate $|\epsilon|$ within the linear regime of induced symmetry breaking, being $\hat P_\phi$ and $\hat X_\phi$ the optimal observables for estimating the amplitude $|\epsilon|$ and phase $\phi$ respectively. We stress the fact that the quantity $r^2_0$ appearing in Eqs.(\ref{Fisher},\ref{FisherPhase}) refers to the number of bosons in the steady state, whose main contribution comes from the incoherent pumping but \textit{not} the probe field, concretely $r_0^2\approx \gamma/\kappa$ in the lowest order. This implies that one can increase the precision in parameter estimation for a fixed driving intensity $\epsilon$ solely by increasing the laser pumping $\gamma$. Additionally, recall that none of the results presented in this work depend on the quantum state of the driving field, as the system steady-state is unique for all of them.
%
%

\section{Criticality as a metrological resource}
\begin{figure}[h!]
  \includegraphics[width=0.45\textwidth]{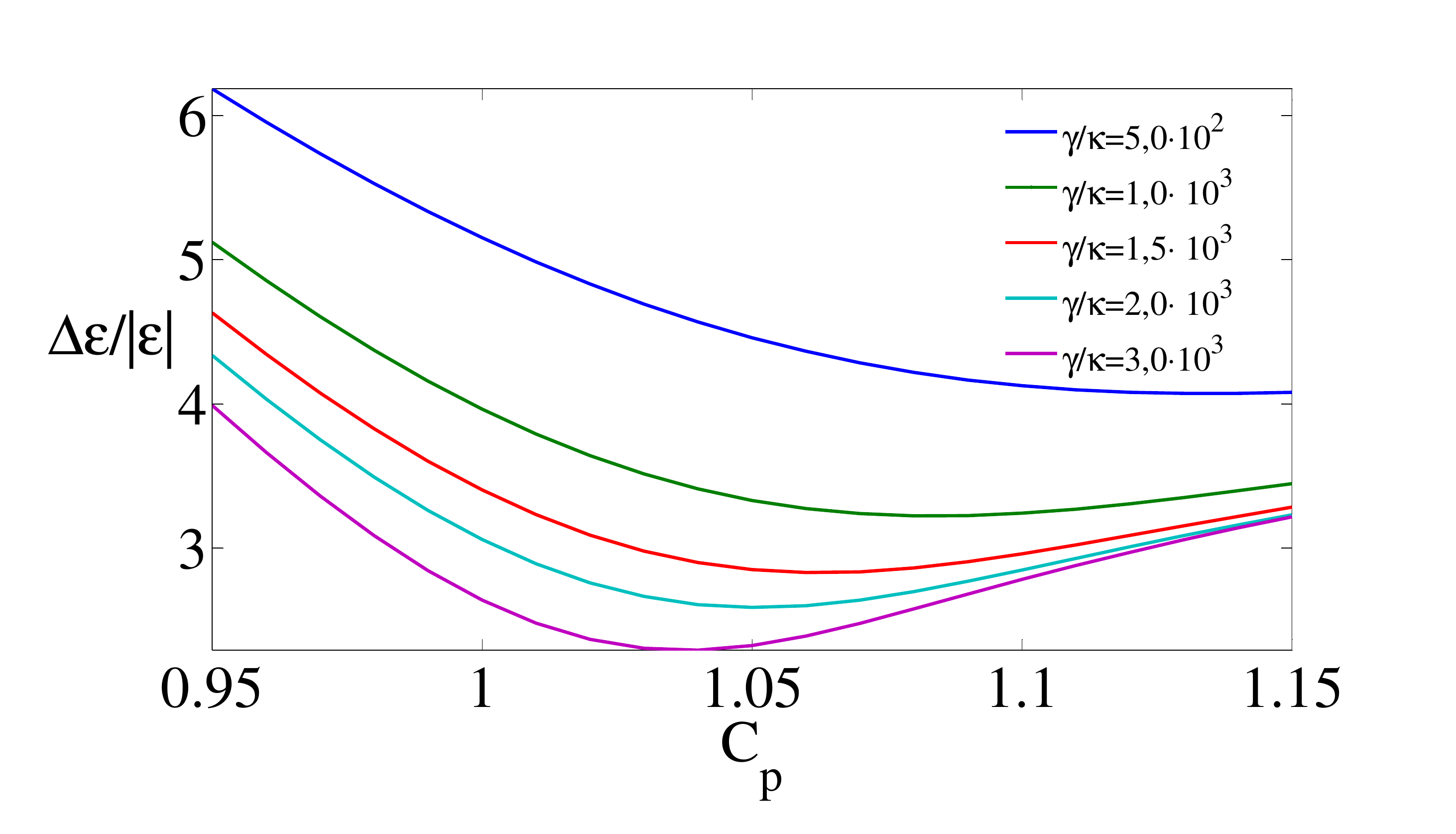}
  \caption{Plot of the relative error ${\Delta|\epsilon|/|\epsilon|}$ by using the average number of bosons as a function of the pump parameter $C_{\rm p}=g^2/\gamma\kappa$ as the thermodynamic limit is approached, $\gamma/\kappa\rightarrow \infty$ ($|\epsilon|=0.1$)}
  \label{fig:fig3} 
\end{figure}

The results obtained in Eqs. (\ref{Fisher}, \ref{FisherPhase}) constitute the maximal metrological capacity of the single qubit laser for estimating $|\epsilon|$ and $\phi$ respectively, as they saturate the Cramer-Rao bound \eqref{CramerRao}. However, estimation by non-optimal observables, like the steady number of bosons $n$, may be experimentally more convenient. Using the analytical results for $n$ and $\Delta n$ (see App.\ref{App:AppendixC}), the expected relative error above threshold for estimating $|\epsilon|$ by means of $n$ is 
\begin{equation}
\frac{\Delta |\epsilon|}{|\epsilon|}=\frac{1}{|\epsilon|}\frac{\Delta n}{\frac{\partial n}{\partial|\epsilon|}}=\frac{C_p\kappa}{g\nu^2}+\frac{g}{2\gamma}\left(\frac{C_p-1}{C_p}\right)+O(|\nu|^2) . \label{nonOptimalEps}
\end{equation}
Eq.\eqref{nonOptimalEps} indicates that the precision increases as we approach the critical point $C_p=1$, at which the precision scales as $\Delta |\epsilon|/|\epsilon|\propto\kappa^{5/2}/(|\epsilon|^2\gamma^{1/2}) 
$\footnote{An exact calculation at the critical point shows that the precise scaling is $\Delta |\epsilon|/|\epsilon|=(\pi/(-4+2\pi))^{1/2}\kappa^{5/2}/(|\epsilon|^2\gamma^{1/2})$}. The ratio between the optimal and non-optimal protocols, $(\Delta |\epsilon|)_{\rm non}/(\Delta |\epsilon|)_{\rm op}\propto \kappa/|\epsilon|$ suggests that both methods give comparable resolutions when $\epsilon\approx \kappa$. Figure \ref{fig:fig3} depicts exact numerical results for $\Delta |\epsilon|/|\epsilon|$, confirming maximal precision around the critical point as the thermodynamic limit is approached. Such limit is reached when $n\rightarrow\infty$ \cite{hwang15}, or equivalently $\gamma/\kappa\rightarrow\infty$. 

The maximal precision given by the critical point manifests a connection between non-equilibrium criticality in dissipative systems and efficient parameter estimation. An analogous result has been already explored for closed systems \cite{Zanardi08}. Physically, it is intuitive to think that the system at the critical point becomes more sensitive to any perturbation, leading to a greater sensor resolution. The potential of criticality for sensing can be exploited in setups where the qubit-boson coupling, $g$, can be controlled with the necessary accuracy to ensure that the system stays at the critical point. This is actually the case in, e.g., single trapped ion phonon lasers, where this coupling is implemented by a laser and its strength modulated by its intensity. Also, in superconducting qubits, qubit-photon coupling terms can be induced and controlled with periodic driving fields \cite{navarrete14prl}. \

A phase estimation by measuring the number of bosons is also possible if we extend the previous setup to arrange an adequate interferometric scheme. Concretely, we add a new reference field term 
$H_{\rm ref}=|\epsilon_0|( a e^{- i \phi_0} + a^{\dag}e^{i\phi_0})$ 
to Eq. \eqref{Hamiltonian}, where we assume that $|\epsilon_0|$, $\phi_0$ are known parameters.
Both the probe field and the reference field must be comparable to observe interference effects, so we shall assume for simplicity that they both have the same amplitude, $|\epsilon_0| = |\epsilon|$. One may treat this new input field as we did in the previous sections, in which case the $P$ function for the steady state will be
\begin{equation}
P(r,\theta)=  \frac{1}{N} \exp{(-\lambda r^4+ \mu r^2 -2\nu' r\sin{(\theta-\phi')} )} \label{PpolarInterf}
\end{equation}
with $\nu'=2\nu\cos[(\phi-\phi_0)/2$ and $\phi'=(\phi+\phi_0)/2$. Comparing Eq.\eqref{PpolarInterf} with Eq.\eqref{Ppolar}, we note that the addition of the reference field to the probe field leads to a total driving field with phase $\phi'$ and amplitude 
$|\epsilon'| = 2 |\epsilon| \cos((\phi-\phi_0)/2)$.
Interference has thus translated the information of $\phi$ into a new phase-dependent amplitude $\epsilon'$, which can be now estimated through measurements of the average boson number with the precision shown in 
Eq. (\ref{nonOptimalEps}). In the lowest order this leads to a precision 
$\Delta \phi \approx C_{\rm p} \kappa / (g \nu^2 \sin(\phi-\phi_0))$, showing that the optimal operating condition is $\phi - \phi_0 = \pi/2$.
%
\section{Possible sources of errors}
One may wonder whether potential sources of error in real experiments could jeopardize our previous results. In the App.\ref{App:AppendixE} we consider three possible sources of error: dephasing of the qubit, heating of the bosonic mode and detuning $\Delta$ between the qubit and the mode. Our calculations show that the detuning is expected to be negligible as long as $\gamma \gg \Delta$, while the dephasing and heating results in a renormalization of the constants $A$ and $B$. In essence, our results are robust to any perturbation that respects the symmetry of the model and the universal scalings of the lasing phase transition.
\section{Physical Implementations}   
Single-qubit photon lasers can be implemented with single atoms \cite{mckeever03} 
or superconducting qubits \cite{Astafiev07,Hauss08,navarrete14prl}. Furthermore, our ideas can be also applied to single-qubit phonon lasers \cite{vahala09,Grudinin10prl}. 
Here, the quantized excitations (phonons) of a trapped ion play the role of the photons in an optical laser, whereas internal electronic levels provide us with a qubit. Our scheme would lead to a the precise measurement of ultra-weak forces of the form 
$H_{\rm f} = F x_0 (a+a^\dagger)$ resonant with the trapping frequency
\cite{biercuk10natnano,schreppler14sci,ivanov15prapp,ozeri16arx}. 
Phonon lasing has actually been already observed in a single trapped ion experiment \cite{vahala09}. All the interactions and techniques required to implement this idea are routinely used in trapped ion experiments, see for example \cite{Leibfried03rmp} for an excellent review on the topic.

To have full control of the parameters involved in our model we will consider a two-ion crystal in which one of the ions acts as a single-atom phonon laser, whereas a second auxiliary ion is used to provide us with a sympathetic cooling mechanism \cite{Barret03pra}. To avoid the requirement of individual addressing of each of the two ions, different species could be used. We assume that ions are weakly coupled by the Coulomb interaction. 
We introduce phonon annihilation operators $a_1$ and $a_2$ associated to quantized vibrations of ions 1 and 2 respectively. The coupling term between the ions takes the form \cite{Porras04prl2,Haze12pra},
\begin{equation}
H_{\rm c} = t_{\rm C} \left( a_1^\dagger a_2 + a_1 a_2^\dagger \right).
\end{equation}
If we consider radial vibrations, then $t_{\rm C} = 2 e^2 /(m d^3 \omega_{\rm T}^2)$, where $d_0$ is the distance between ions, $m$ refers to the ion's mass and $\omega_{\rm T}$ is the trapping frequency.

Let us consider now the first ion's quantum dynamics. To make the connection with trapped ion physics clearer, we will work in a spin basis where the role of states $|e\rangle$  and $|g\rangle$ is interchanged with respect to the discussion in the main text. 
In our trapped ion scheme, spin pumping will be induced by the radiative decay from an excited state $|e\rangle$ to the ground state $|g\rangle$, whereas a spin-phonon coupling of the form $(\sigma^+ a^\dagger + \sigma^- a)$ will be induced. This is described by the following Liouvillian,

\begin{equation}
{\cal L}_{1}(\rho) = - i [H_{1}, \rho] + {\cal L}_{\{ \sigma_1^-,\gamma \}} (\rho).
\label{L1}
\end{equation}
The Hamiltonian acting on ion $1$ includes a blue-sideband coupling between the internal state of the ion and the local vibrational mode as well as the coupling to the external force that we aim to measure,
\begin{equation}
H_1 = g (\sigma^+_1 a^\dagger_1 + \sigma^-_1 a_1) + \epsilon (a^\dagger_1 + a_1).
\end{equation}
We have introduced ladder operators, $\sigma^+_1$, $\sigma^-_1$, associated to the internal state of ion 1. The blue side-band term can be induced by lasers with frequency $\omega_{\rm L} = \omega_0 + \omega_{\rm T}$, where $\omega_0$ is the frequency of the internal state transition \cite{Leibfried03rmp}. Finally, the last term of Eq (\ref{L1}) is simply the radiative decay of the excited state \cite{Leibfried03rmp}. To ensure that the dynamics of the ion is constrained to only two levels, one could simply choose $|g\rangle$ and $|e\rangle$ as the two levels of a cycling transition.

The only missing element is a cooling mechanism acting on ion 1. For this we will use ion 2 to provide us with a cooling medium by an effect known as sympathetic cooling. For this we assume that ion 2 is being continuously laser cooled with a rate $\kappa_{2}$, 
\begin{equation}
{\cal L}_{2} =  {\cal L}_{\{ a_2,\kappa \}} (\rho).
\end{equation}
If the Coulomb coupling is small relative to the cooling rate ($t_{\rm C} \ll \kappa_2$), we can adiabatically eliminate ion 2 and obtain an effective cooling term for ion 1, with cooling rate $\kappa_{\rm eff} = t_{\rm C}^2/\kappa_2$. The reduced density matrix for ion 1, $\rho_1$, is thus subjected to the following quantum dynamics,
\begin{equation}
\dot{\rho}_1 = {\cal L}_1(\rho_1) +
 {\cal L}_{\{ \kappa_{\rm eff}, a_1 \}}(\rho_1) .
\end{equation}

Our scheme is a phononic version of the single-qubit laser described in the main text. To assess the sensitivity of such a device in the measurement of external forces, we consider now some typical values for cooling rates and vibrational couplings. We focus on the optimal measurement protocol, which would imply measuring the quadrature, $\hat{P}_{\phi}$, defined above Eq. (6) of the main text. Quadratures of vibrational operators can be efficiently measured by coupling phonon observables to the ion's internal state and detecting the emitted fluorescence (see for example \cite{Leibfried03rmp}). By using our calculation of the error as estimated from the QFI we get
\begin{equation}
\Delta \epsilon = \frac{1}{\sqrt{F_Q[\rho_{|\epsilon|}]}} = \frac{A}{\sqrt{2} r_0} .
\end{equation}
To estimate $A$, we express it like $A = g^2/\gamma = C_{\rm p} \kappa_{\rm eff} \approx \kappa_{\rm eff}$, where we have assumed that we work in a regime with cooperativity parameter $C_{\rm p} \approx 1$. 

Our scheme can be applied to measure ultra-weak forces. The relation between the driving strength $|\epsilon|$ and the applied external force, $F$, is $|\epsilon| = F x_0$, where
\begin{equation}
x_0 = \frac{1}{\sqrt{2 m \omega_{\rm T}}},
\end{equation}
is the size of the vibrational ground state. Our final expression for the force sensitivity reads (in standard units including $\hbar$),
\begin{equation}
\Delta F \approx \frac{\hbar \kappa_{\rm eff}}{\sqrt{2 n_{\rm ph}} x_0},
\label{F.prec}
\end{equation}
where we have used the fact that the number of phonons, $n_{\rm ph} \approx r_0^2$. To get an estimate of the precision with which an ultra-weak force could be measured, we consider that ion 1 is $^{40}$Ca$^+$ and $\omega_{\rm T}/(2 \pi) =$ 10 MHz, which yields $x_0 =$ 3.5 nm. Other typical values are $t_{\rm C}/(2 \pi) = 4$ kHz \cite{Haze12pra} and $k_2/(2 \pi) = 40$ kHz, leading to $\kappa_{\rm eff}/(2 \pi) = 0.4$ kHz. With those values we get
\begin{equation}
\Delta F \approx 53 \ {\rm yN}  / \sqrt{n_{\rm ph}} ,
\label{F.prec}
\end{equation}
By increasing the number of phonons in the lasing regime to values such as ${n_{\rm ph} = 2 \times 10^3}$, one could obtain precisions $\Delta F \approx 1.2$ yN, well within the yocto-Newton regime and beyond the precision of results reported in experiments \cite{biercuk10natnano}. 

Large phonon numbers are in principle not difficult to get in a trapped ion phonon laser. For example, taking into account typical values of $\gamma/ (2 \pi) = 20$ MHz our Eq.(\ref{nph}) yields the value ${n_{\rm ph} = 2 \times 10^3}$ with a side-band coupling $g/(2 \pi) = 66.5$ kHz, well within the state-of-the-art \cite{Leibfried03rmp}.

A limiting factor could be the presence of motional heating, $\kappa_{\rm h}$. However, heating rates in linear Paul traps can be as low as 0.1 vibrational quanta per ms, which translates into $\kappa_{\rm h}/(2 \pi) = 0.008$ kHz \cite{Turchette00pra}. Under those conditions, $\kappa_{\rm eff} \gg \kappa_{\rm h}$ and the effect of heating could be neglected or incorporated into minor corrections to the trapped ion sensor (see section VI).
\section{Acknowledgments} Funded by the People Programme (Marie Curie
Actions) of the EU’s Seventh Framework Programme under REA Grant Agreement no: PCIG14-GA-2013-630955. We thank Jacob Dunningham and Pedro Nevado for fruitful discussions.


\appendix

\section{Adiabatic elimination}\label{App:AppendixA}

Here we shall derive the effective quartic master equation claimed in \eqref{finalrhof}, as a result of the adiabatic elimination of the fast spin variable. Firstly, we shall trace over the spin degree of freedom from the master equation for the single-qubit laser,
\begin{equation}
\dot{\rho}=-i [H,\rho]+\mathcal{L}_{\{\sigma^+ , \gamma\}}(\rho) + 
\mathcal{L}_{\{a , \kappa\}}(\rho),
\label{Liouvillian2}
\end{equation}
thereby obtaining an equation for the reduced density matrix of the bosonic field 
$\dot{\rho_f} = {\rm Tr}_{\rm qubit}{\{\mathcal{L}(\rho)\}}$.
Namely, this equation reads
\begin{multline}
\dot{\rho}_f=-i g(a \rho_{ge} +a^{\dag} \rho_{eg} -\rho_{ge}a -\rho_{eg}a^{\dag}) - \\
	-i (\epsilon \adag \rho_f +\epsilon^* a \rho_f -\epsilon\rho_f\adag -\epsilon^*\rho_f\adag)+ \\
	+\kappa (2a \rho_f \adag - \adag a \rho_f -\rho_f\adag a) , \label{rhof}	
\end{multline}
where we introduced the notation $\rho_{ge}=\langle g| \rho|e \rangle=\rho_{eg}^{\dag} $ and $\epsilon=|\epsilon|e^{i\phi}$. To obtain a closed equation for the reduced density matrix $\rho_f$, we have to eliminate the operators $\rho_{ge},\rho_{eg}$ from Eq. \eqref{rhof}. We obtain the corresponding equations of motion for these operators using the original master equation,
\begin{equation}
\dot{\rho}_{ge}=-i g(\adag\rho_{ee} -\rho_{gg}\adag) - \gamma \rho_{ge} ,\label{rhoge}
\end{equation}
where we have neglected the contributions from $\kappa$ and $\epsilon$ in comparison with $\gamma$. In the limit  $\gamma \gg \kappa, g, |\epsilon|$, we can adiabatically eliminate the operators $\rho_{ge}$ and $\rho_{eg}$ from \eqref{rhof} by taking $\dot{\rho}_{ge}\approx 0$ in Eq.\eqref{rhoge}  and substituting their steady-state solutions,
\begin{equation}
\rho_{ge}=-i \frac{g}{\gamma}(\adag \rho_{ee}-\rho_{gg} \adag) . \label{rhoeg}
\end{equation}
As the resulting equation still depends on the operators $\rho_{gg}$ and $\rho_{ee}$, we make use of the single-qubit master equation to obtain the equations of motions of these operators,
\begin{align}
\dot{\rho}_{ee} &=-i g(a \rho_{ge}-\rho_{eg} \adag) +2\gamma \rho_{gg} \label{rhoeedot}\\ 
\dot{\rho}_{gg} &=-i g(\adag \rho_{eg}-\rho_{ge} a) -2\gamma \rho_{gg} \label{rhoggdot}
\end{align}
where we again neglect terms with $\kappa$ and $\epsilon$. One may now obtain a perturbative solution to the steady-states of Eqs.\eqref{rhoeedot}\eqref{rhoggdot} in terms of the field density matrix $\rho_f$. To do so, let us adiabatically eliminate $\rho_{gg}$ by taking $\dot{\rho}_{gg}\approx 0$ in Eq.\eqref{rhoggdot}, yielding
\begin{equation}
\rho_{gg}= -\frac{ig}{2\gamma}(\adag \rho_{eg}-\rho_{ge} a)  \label{rhogg}
=\frac{g^2}{2\gamma^2}(2\adag \rho_{ee}a-\adag a \rho_{gg}-\rho_{gg}\adag a).
\end{equation}
In a first order approximation, the ground state population is negligible due to the fast pumping of the atoms ($\gamma \gg 1$). Therefore, we expect to find $\rho_{gg}\approx 0$ and $\rho_{ee}=\rho-\rho_{gg}\approx \rho_f$ in first order. A second order correction is achieved by inserting this first order approximation into Eq.\eqref{rhogg}, hence
\begin{align}
\rho_{gg}&= \frac{g^2}{\gamma^2}\adag \rho_f a \label{rhogg2}\\ 
\rho_{ee}&=\rho_f-\rho_{gg}=\rho_f- \frac{g^2}{\gamma^2}\adag \rho_f a \label{rhoee}.
\end{align}
One can finally insert Eqs.\eqref{rhogg2}\eqref{rhoee} into Eq.\eqref{rhof} to arrive at the desired closed equation for $\rho_f$,
\begin{multline}
	\dot{\rho}_f= -i (\epsilon \adag \rho_f +\epsilon^* a \rho_f -\epsilon\rho_f\adag -\epsilon^*\rho_f a)+ \\ 
		+\frac{g^2}{\gamma}( 2 \adag \rho_f a -a\adag \rho_f- \rho_f a\adag) + \\
		+\frac{2g^4}{\gamma^3}(  a\adag \rho_f a \adag-\adag^2 \rho_f a^2) + \\
		+\kappa (2a \rho_f \adag - \adag a \rho_f -\rho_f\adag a) . \label{adiabaticEquation}
\end{multline}
The second term in the r.h.s. of Eq.\eqref{adiabaticEquation} accounts for the single photon emission by the excited qubit (linear gain), while the third represents the contribution of two cycles of emission and re-excitation (gain saturation). Eq.\eqref{adiabaticEquation} can be cast in Lindblad form as presented Eq.\eqref{finalrhof}. A few brief remarks are worth mentioning about the single-qubit laser physics. Using Eq.\eqref{adiabaticEquation} and setting $\epsilon=0$, we can easily derive an equation for the diagonal elements $\rho_{nn}$, namely
\begin{multline}
	\dot{\rho}_{nn}= -(2A-B(n+1))(n+1)\rho_{nn} \\
	+2An\rho_{n-1,n-1} -Bn(n-1) \rho_{n-2,n-2} \\
	-2C n \rho_{nn}+2C (n+1) \rho_{n+1,n+1}, \label{rhonn}
\end{multline}
where we defined the coefficients $A=g^2/\gamma$, $B=2g^4/\gamma^3$ and $C=\kappa$. In contrast to the classic \textit{Scully-Lamb} treatment of the four-level laser\cite{Sargent74}, no detail balance solution can be found to Eq.\eqref{rhonn}. The rate equation for the average photon number $\langle n \rangle= \sum{n\rho_{nn}}$ can also be derived from \eqref{rhonn},
\begin{equation}
\langle \dot{n} \rangle=2(A-C)\langle n\rangle+2A-B(2\langle n^2 \rangle +5\langle n \rangle +5). \label{raten}
\end{equation}
According to Eq. \eqref{raten}, there will be an initial exponential increase in the mean photon number if $A>C$, hence $A=C$ is the threshold condition for the laser phase. This agrees with the prediction of a mean field treatment to this problem, in which the lasing phase is found when the pump parameter $C_{\rm p}\equiv g^2/\gamma\kappa$ satisfies $C_{\rm p}>1$ \cite{BreuerPet}. \\
We now address the conditions of validity of the adiabatic elimination. Using Eq. (\ref{rhogg2}), the condition $\rho_{gg}\approx 0$ is translated into 
\begin{equation}
\langle \sigma^+ \sigma^- \rangle = \rm Tr\{ \rho_{gg} \} = 
\left( \frac{g}{\gamma} \right)^2 \left(1 + n \right) \ll 1.
\end{equation}
Below threshold ($C_{\rm p} < 1$), this is satisfied as long as $g/\gamma \ll 1$. Above threshold ($C_{\rm p} > 1$), we can estimate $n = (A-C)/B$ (see App. \ref{App:AppendixC}), which leads to
\begin{equation}
\langle \sigma^+ \sigma^- \rangle \approx \frac{1}{2} \frac{C_{\rm p} - 1}{C_{\rm p}}\ll 1 .
\end{equation}
Consequently, for the adiabatic elimination to be self-consistent above threshold we must require $C_{\rm p} \gtrsim 1$.
%
\section{Laplace's method} \label{App:AppendixC}

We shall calculate different observables associated to the $P(\alpha,\alpha^*)$ function obtained in \eqref{Ppolar}, corresponding to the steady-state solution of the Fokker-Planck equation \eqref{PFokker}. To do so, it is first necessary to compute the normalization constant $N$ given in \eqref{Ppolar}. This can be approximately integrated using the Laplace's method, which is helpful for integrals of the form
\begin{equation}
I(s)=\int_{-\infty}^\infty f(x) e^{sg(x)} dx \approx \sqrt{\frac{2\pi}{sg''(x_0)}} f(x_0)e^{sg(x_0)} , \label{Laplace}
\end{equation}
in which $x_0$ stands for the global maximum of $g(x)$, $g''(x_0)$ represents its second derivative evaluated at $x_0$, $f(x)$ varies slowly around $x_0$ and is independent of the parameter $s$. In our case, $N$ has the form
\begin{equation}
N=\int_{0}^\infty\int_{0}^{2\pi} rd\theta dr e^{(-\lambda r^4+ \mu r^2 -2\nu r\sin{(\theta-\phi} ))}.
\end{equation}
Integrating over $\theta$ gives
\begin{equation}
N=2\pi\int_{0}^\infty r dr I_0(2\nu r)e^{(-\lambda r^4+ \mu r^2 )} .
\end{equation}
where $I_n$ are the modified Bessel functions of the first kind. 
Above threshold, where $\mu\gg \lambda$, the normalization constant $N$ is approximated by Eq.\eqref{Laplace} as
\begin{equation}
N\approx \sqrt{\frac{\pi^3}{\lambda}}I_0(2\nu r_0) \exp{\left(\frac{\mu^2}{4\lambda}\right)},
\end{equation}
where $r_0=\sqrt{\mu/2\lambda}$. The laser field quadrature $\langle \hat{P}_\phi \rangle=\langle i(ae^{-i\phi}-a^{\dag}e^{i\phi}) \rangle$ may now be computed by taking the parametric derivative $\langle  \hat{P}_\phi  \rangle=N^{-1}(\partial N/\partial \nu)$, which gives
\begin{equation}
\langle \hat{P}_\phi  \rangle=2r_0\frac{I_1(2\nu r_0)}{I_0(2\nu r_0)}. \label{fieldAlpha}
\end{equation}
Assuming that $|\epsilon| \ll 1$, one may expand \eqref{fieldAlpha} in Taylor series as
\begin{equation}
\langle \hat{P}_\phi \rangle\approx 2\nu r_0^2-{\nu^3 r_0^4}+O(\nu^4), 
\end{equation}
which in first order indicates a linear dependence in  $|\epsilon|$ as claimed Eq.\eqref{quadrature}. To compute the uncertainty of $\langle \hat{P}_\phi \rangle$ a second derivative is required, specifically $\Delta^2\hat{P}_\phi=\langle\hat{P}^2_\phi\rangle-(\langle\hat{P}_\phi\rangle)^2= (\partial^2 N/\partial \nu^2)/N-((\partial N/\partial \nu)/N)^2$, the result of which reads

\begin{equation}
 \Delta^2\hat{P}_\phi= 2r_0^2\left( 1+ \frac{I_2(2\nu r_0)}{I_0(2\nu r_0)}- 2\left(\frac{I_1(2\nu r_0)}{I_0(2\nu r_0)}\right)^2  \right). \label{uncertQuad}
\end{equation}
When $|\epsilon| \ll 1$, a Taylor expansion of \eqref{uncertQuad} gives
\begin{equation}
 \Delta^2\hat{P}_\phi=  2r_0^2\left(1-\frac{3\nu^2r_0^2}{2} +O(\nu^4) \right).
\end{equation}
On the other hand, the parametric derivatives with respect to $\mu$ can be related to the average number of bosons and its uncertainty. First, the average number of bosons $\langle n \rangle=\langle r^2 \rangle$ is given by  $\langle n  \rangle=N^{-1}(\partial N/\partial \mu)$, yielding
\begin{equation}
\langle n\rangle=  r_0^2+\frac{\nu r_0}{\mu}\frac{I_1(2\nu r_0)}{I_0(2\nu r_0)}=r_0^2+\frac{\nu}{2\mu}\langle \hat{P}_\phi\rangle . \label{averageN}
\end{equation}
If $|\epsilon| \ll 1$, Eq.\eqref{averageN} is approximated by 
\begin{equation}
\langle n\rangle=  r_0^2+\frac{r_0^2\nu^2}{\mu}-O(\nu^4) .
\end{equation}
\begin{figure}[h!]
	\includegraphics[width=0.45\textwidth]{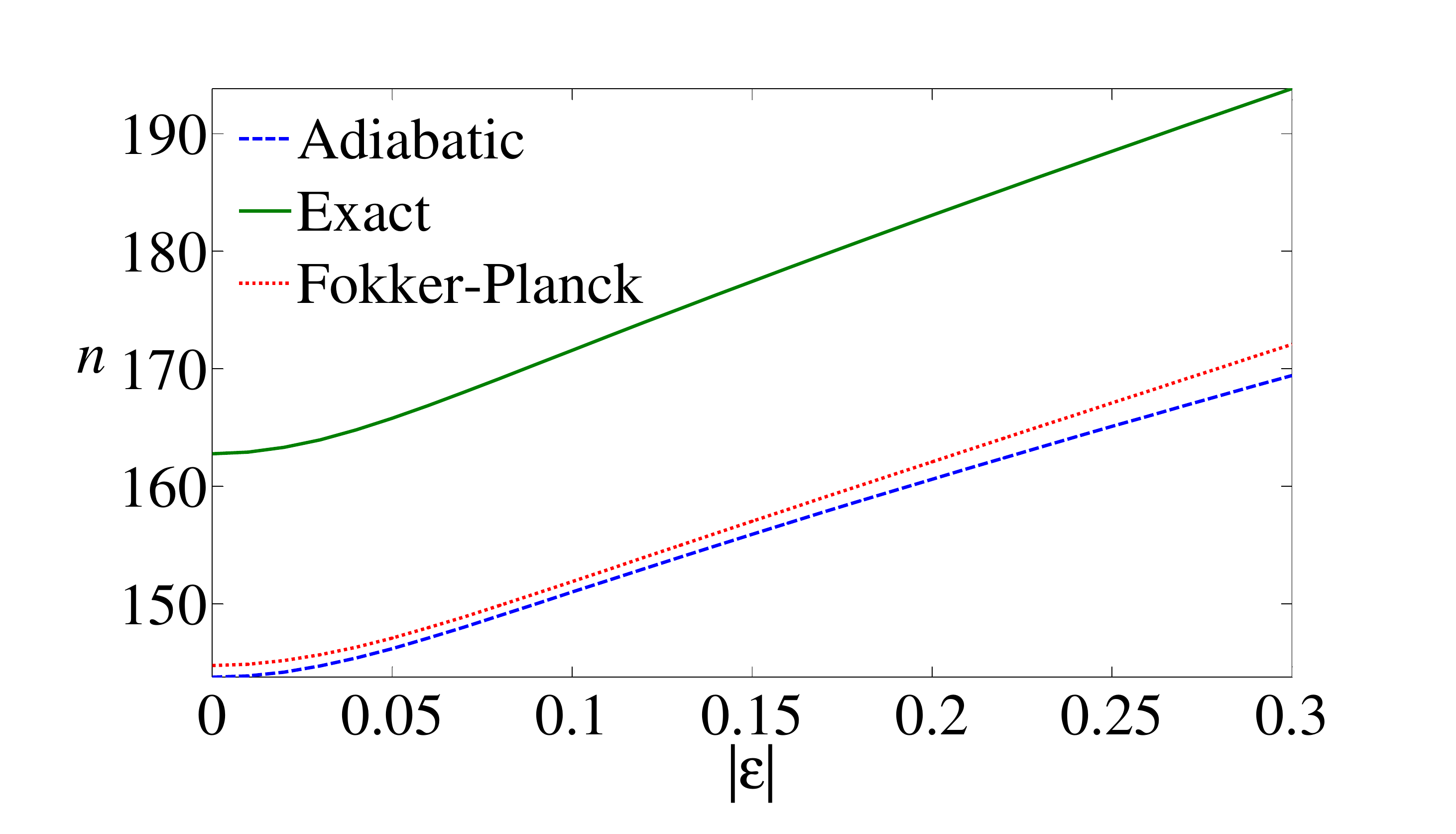} 
  \caption{Plot of the averaged number of bosons as a function of the amplitude $|\epsilon|$, showing the comparison of exact calculations (solid line), the adiabatic equation \eqref{adiabaticEquation} (dashed line) and analytical results by the Laplace's method (dotted line). Parameters: $g=58,\gamma=3000,\kappa=1$}
\end{figure}
The corresponding uncertainty can be computed as $\Delta^2 n=\langle n^2\rangle-(\langle n\rangle)^2= (\partial^2 N/\partial \mu^2)/N-((\partial N/\partial \mu)/N)^2$. The final result of such calculation gives
\begin{multline}
 \Delta^2 n=  \frac{2\mu+\nu^2}{4\lambda\mu}-\frac{\nu^2}{2\lambda\mu}\left(\frac{I_1(2\nu r_0)}{I_0(2\nu r_0)}\right)^2- \\
-\frac{r_0\nu}{2\mu^2}\frac{I_1(2\nu r_0)}{I_0(2\nu r_0)} +\frac{\nu^2}{4\lambda\mu}\frac{I_2(2\nu r_0)}{I_0(2\nu r_0)}.\label{uncertN}
\end{multline}
If $|\epsilon| \ll 1$, $\Delta^2 n$ is approximated by 
\begin{equation}
 \Delta^2 n=  \frac{1}{2\lambda}-\frac{\nu^4}{8\lambda^2}+O(\nu^8) .
\end{equation}
Finally, the averaged field quadrature $\langle \hat{X}_{\phi} \rangle=\langle (ae^{-i\phi}+a^{\dag}e^{i\phi}) \rangle$ and its uncertainty are needed to compute the Quantum Fisher information. The former is directly given by the parametric derivative $\langle \hat{X}_{\phi} \rangle=(\partial N/\partial \phi)/N=0$. From Eq.\eqref{Ppolar} one can show that the relation  $(\partial^2 N/\partial \phi^2)/N=-\nu\langle \hat{P}_{\phi} \rangle+\nu^2\langle \hat{X}^2_{\phi} \rangle$ holds, allowing us to compute the field uncertainty  $\Delta^2 \hat{X}_{\phi} $ analytically as
\begin{equation}
 \Delta^2\hat{X}_{\phi}=\langle \hat{X}^2_{\phi} \rangle=\nu^{-1}\langle \hat{P}_{\phi}\rangle,
\end{equation}
since $\partial N/\partial \phi=0$.
\section{Sources of error} \label{App:AppendixE}
In this section we discuss in more detail how possible sources of error could affect the ideal dynamics as presented in Eq.\eqref{Liouvillian2}. On the one hand, we consider two possible noise terms for the dissipation: a nonradiative dephasing process of the qubit at rate $\gamma_{\rm dep}$ plus heating of the bosonic mode at rate $\kappa_{\rm h}$. These terms can be modeled as
\begin{equation}
 (\dot{\rho})_{\rm err}=-\frac{\gamma_{\rm dep}}{2}(\sigma^z\rho\sigma^z -\rho)+\kappa_{\rm h}( 2 \adag \rho_f a -a\adag \rho_f- \rho_f a\adag),
\end{equation}
which have to be added to the general master equation \eqref{Liouvillian2}. The dephasing term changes the equation \eqref{rhoge} for $\rho_{eg}$ just by a renormalization of the pumping $\gamma'=\gamma+ \gamma_{\rm dep}/2$ whereas leaving the equations \eqref{rhoggdot}\eqref{rhoeedot} intact. Hence, equations \eqref{rhoeg}\eqref{rhogg2} are modified as
\begin{align}
\rho_{ge}&=-i \frac{g}{\gamma'}(\adag \rho_{ee}-\rho_{gg} \adag)\\ 
\rho_{gg}&=\frac{g^2}{\gamma\gamma'}\adag \rho_f a .
\end{align}
The heating term has the same form as the second term in Eq.\eqref{rhof}. As a result, the effect of these process turns out to be a renormalization of the coefficients $A,B$, such that
\begin{equation}
 A'=\frac{g^2}{\gamma'}+\kappa_{\rm h}, \quad B'=\frac{g^4}{(\gamma')^2\gamma}\approx \frac{g^4}{\gamma^3+\gamma^2\gamma_{\rm dep}}.
\end{equation}
On the other hand, we also consider a possible detuning between the mode frequency $\omega$ and the qubit frequency $\delta$. In an interaction picture rotation at the mode frequency, this effect is included as a new term in the Hamiltonian as follows
\begin{equation}
H = H_{\rm JC}+H_{\rm d}+\frac{\Delta}{2}\sigma^z , 
\end{equation}
where $\Delta=\delta-\omega$ is the detuning. This term alters equations \eqref{rhoeg}\eqref{rhogg2} as follows
\begin{align}
\rho_{ge}&=-i \frac{g}{\gamma-i\Delta}(\adag \rho_{ee}-\rho_{gg} \adag)\\ 
\rho_{gg}&=\frac{g^2}{\gamma^2+\Delta^2}\adag \rho_f a .
\end{align}
Consequently, equation \eqref{adiabaticEquation} is modified by adding a new term with a prefactor $g^2\Delta/(\gamma^2+\Delta)$ that can be safely neglected, and a renormalization of the coefficients $A,B$, such that
\begin{equation}
 A'=\frac{g^2}{\gamma(1+(\frac{\Delta}{\gamma})^2)}, \quad  B'=\frac{g^4}{\gamma^3(1+(\frac{\Delta}{\gamma})^2)}.
\end{equation}
As we are in the strong pumping regime  $\gamma \gg 
\Delta$, the effect of a possible small detuning is expected to be negligible. In conclusion, we observe that the possible sources of error considered, i.e. dephasing, heating and detuning, result in a renormalization of the constants defined in the ideal case, but they are not expected to jeopardize the sensing process or the performance in a significant way. 

\bibliography{biblio}

\end{document}